\documentclass[prc]{revtex4}
\usepackage{graphicx,amsmath}
\begin{document}
\title{Statistical and Dynamic Models of Charge Balance Functions}
\author{Sen Cheng}
\altaffiliation{Current address: 
Sloan-Swartz Center, University of California at San Francisco\\
513 Parnassus Ave., Box 0444, San Francisco, CA 94143-0444, USA}
\affiliation{Department of Physics and Astronomy, and National Superconducting
Cyclotron Laboratory\\ Michigan State University, East Lansing Michigan, 48824}
\author{Charles Gale}
\affiliation{Physics Department, McGill University, Montr\'eal, Quebec, 
H3A 2T8, Canada}
\author{Sangyong Jeon}
\affiliation{Physics Department, McGill University, Montr\'eal, Quebec, 
H3A 2T8, Canada}
\author{Silvio Petriconi}
\affiliation{Department of Physics and Astronomy, and National Superconducting
Cyclotron Laboratory\\ Michigan State University, East Lansing Michigan, 48824}
\author{Scott Pratt}
\affiliation{Department of Physics and Astronomy, and National Superconducting
Cyclotron Laboratory\\ Michigan State University, East Lansing Michigan, 48824}
\author{Michael Skoby}
\altaffiliation{Current address: Department of Physics, Purdue
University, 525 Northwestern Avenue, West Lafayette, IN 47907} 
\affiliation{Department of Physics and Astronomy, and National Superconducting
Cyclotron Laboratory\\ Michigan State University, East Lansing Michigan, 48824}
\author{Vasile Topor Pop}
\affiliation{Physics Department, McGill University, Montr\'eal, Quebec, 
H3A 2T8, Canada}
\author{Qing-Hui Zhang}
\affiliation{Physics Department, McGill University, Montr\'eal, Quebec, 
H3A 2T8, Canada}
\date{\today}

\begin{abstract}
Charge balance functions, which identify balancing particle-antiparticle pairs
on a statistical basis, have been shown to be sensitive to whether
hadronization is delayed by several fm/c in relativistic heavy ion
collisions. Results from two classes of models are presented here, microscopic
hadronic models and thermal models. The microscopic models give results which
are contrary to recently published $\pi^+\pi^-$ balance functions from the STAR
collaboration, whereas the thermal model roughly reproduce the experimental
results. This suggests that charge conservation is local at breakup, which is
in line with expectations for a delayed hadronization. Predictions are also
presented for balance functions binned as a function of $Q_{\rm inv}$.
\end{abstract}

\maketitle

\section{Introduction}
\label{sec:intro}
Energy densities of near 10 GeV/fm$^3$ covering volumes of several hundred
fm$^3$ are attained in Au+Au collisions at the Relativistic Heavy Ion Collider
(RHIC). Given that the volume of a typical hadron is approximately one fm$^3$,
this far exceeds the energy density of a typical hadron. Therefore, quark and
gluon degrees of freedom are expected to provide a meaningful basis for
describing the microscopic motion for several fm/c, until the matter expands
and cools to a point where hadronic degrees of freedom again become
appropriate.

The conversion from partonic to hadronic degrees of freedom should be
accompanied by a large increase in the number of quark-antiquark pairs as the
entropy stored in gluons and quarks is converted to hadrons, each of which has
at least two quarks. These newly created charges are significantly more
correlated to their balancing anti-charges than those charge pairs created in
the early stages of the collision. In fact, these newly created pairs should be
more tightly correlated than pairs from $pp$ collisions. In $pp$ collisions,
most pairs are created within the first one fm/c in the decay of color flux
tubes, or strings, which involves the separation of the balancing quarks
through tunneling. If the quarks separate by a distance of 0.5 fm at a time of
0.5 fm/c, they find themselves in regions where the collective rapidities
differ by one unit since the velocity gradient along the beam axis is
approximately $1/\tau$. In contrast, a quark that is produced at 5 fm/c, and is
0.5 fm from its balancing partner, is separated by only a tenth of a unit of
rapidity. Thus, a measurement of the relative separation of balancing charges
would yield invaluable information concerning whether hadronization was delayed
beyond the characteristic hadronic time scale of one fm/c \cite{bassdanpratt}.

Charge balance functions provide the means to identify balancing charges on a
statistical basis through a like-sign subtraction. The balance function is
defined as a conditional distribution,
\begin{equation}
\label{eq:balancedef}
B(P_2|P_1)\equiv\frac{1}{2}\left\{
\frac{N_{+-}(P_1,P_2)-N_{++}(P_1,P_2)}{N_+(P_1)}
+\frac{N_{-+}(P_1,P_2)-N_{--}(P_1,P_2)}{N_-(P_1)}\right\},
\end{equation}
where the $+/-$ indices refer to particles or anti-particles. Expressed in
words, the balance function measures the probability of observing an extra
particle of the opposite sign with momentum $P_2$ given the observation of the
first particle with momentum $P_1$. Typically, $P_1$ will refer to a particle
observed anywhere in the detector, and $P_2$ will refer to either the relative
rapidity $\Delta y$ or the relative momentum $Q_{\rm inv}$. In such a case, the
balance function is then labeled by only one variable, e.g., $B(\Delta y)$.

\begin{figure}
\centerline{\includegraphics[width=0.45\textwidth]{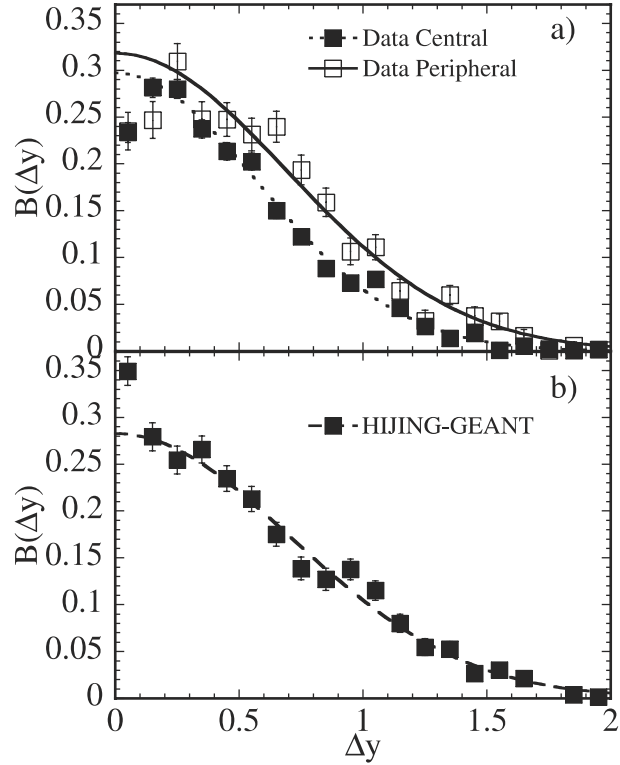}}
\caption{\label{fig:star_pipi}
Balance functions for charged pion pairs as measured by the STAR
collaboration. For each charged pion, there is an enhanced probability of
finding an extra charged pion of the opposite sign within a unit of
rapidity. The peripheral data are well reproduced by the HIJING model which can
be considered as being caused by overlapping independent $pp$ collisions.}
\end{figure}

\begin{figure}
\centerline{\includegraphics[width=0.45\textwidth]{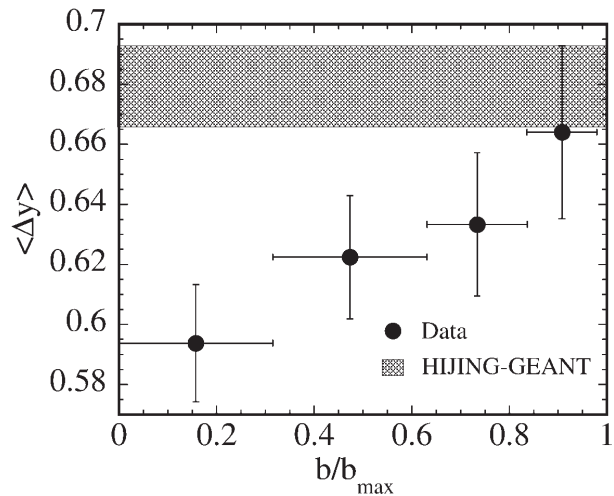}}
\caption{\label{fig:star_widths}
The mean widths of $\pi^+\pi^-$ balance functions as measured by
STAR. For small impact parameters, $b$ the balance function becomes narrower,
qualitatively consistent with expectations for a quark-gluon plasma. For
peripheral collisions, the width matches predictions from HIJING.}
\end{figure}

The STAR collaboration has recently published the first measurement of charge
balance functions in relativistic heavy ions \cite{starbalance}. Measurements were
performed for all charged particles as a function of the relative
pseudo-rapidity and as a function of relative rapidity for identified
pions. Indeed, the presence of the extra balancing charges was apparent and
statistics were sufficient to make distributions as a function of the relative
rapidity as illustrated in Fig. \ref{fig:star_pipi}. Remarkably, the width of
the balance function in $\Delta y$ decreased with the centrality of the
collision, qualitatively consistent with expectations for an increasingly
delayed hadronization as shown in Fig. \ref{fig:star_widths}.

The purpose of this paper is to compare the behavior seen by STAR with two
classes of calculations. First, we compare to two microscopic hadronic
simulations, RQMD\cite{rqmd} and HIJING/GROMIT\cite{hijing,gromit}. Both
approaches involve creating the bulk of hadrons instantaneously according to
$pp$ phenomenology with the subsequent rescattering modeled by a hadronic
cascade. As will be seen in Section
\ref{sec:gromit} these models give qualitatively different behavior
than what was observed by STAR, i.e., the width of the balance functions
increases with the centrality of the collision rather than decreases. Sections
\ref{sec:thermal}, \ref{sec:interdomain} and \ref{sec:thermalresults} present 
the formulation and results of a thermal model which inserts charge
conservation into the blast-wave model which has been successful in describing
spectra. In this model an ensemble of particles is simulated from a single
thermal domain in such a way that the net baryon number, electric charge and
strangeness are zero. These particles then have their momenta and space-time
coordinates re-assigned according to a blast-wave prescription. The effect of
inter-domain interactions, which were shown to be non-negligible in
\cite{balancedistortion}, are also taken into account and described in Sec. \ref{sec:interdomain}. The resulting balance
functions are in remarkable agreement with the STAR measurement provided the
domain is constrained to being highly localized in coordinate space, as would
be expected in the delayed-hadronization scenario. The final section presents
an interpretation of this comparison and presents a discussion of the prospects
for making similar comparisons with $K^+K^-$ or $p\bar{p}$ balance functions.

\section{Balance functions from microscopic hadronic simulations}
\label{sec:gromit}

One fm/c after the initial collision the energy densities achieved in Au+Au
collisions at RHIC are in the neighborhood of 5-10 GeV/fm$^3$
\cite{rhicbjorkenestimate}. In hadronic simulations, hadrons are generated from
overlapping nucleon-nucleon simulations with the creation times typically being
less than a fm/c. Thus, such calculations simulate the evolution of the
collision by assuming hadrons are formed and that they propagate in an
environment that should preclude their very existence due to the high energy
density. Despite the inherent inconsistency in these approaches, such
calculations are important as they provide a baseline from which one can
understand the degree to which novel degrees of freedom and collective
phenomena alter the final state. Here, we present results from several such
models. The first is RQMD \cite{rqmd} which has been the work horse of such
models over the the last decade. The second approach represents a merging of
HIJING and a hadronic cascade GROMIT, which, like RQMD, models the final-state
interactions through the scattering of the hadrons. This second approach is
less sophisticated than RQMD, but has an advantage in speed, allowing the
compilation of 20 thousand central events. Comparisons with many of the same
models have been performed for charge fluctuations \cite{mcgillchargefluc},
which are intimately related to balance functions \cite{jeonpratt}. 

For the purposes of this paper, the important aspect of the simulations is that
they provide a model where the various charges are created early, experience
numerous interactions, and are eventually emitted into the final state. In both
RQMD and HIJING/GROMIT, the charges are created principally by the initial
fragmentation of strings. Like other string-based microscopic models
\cite{venus,hsd,art,nexus,rqmd,urqmd,ampt,jpciae,qgsm}, these models
incorporate a formation time of the scale of one fm/c. These approaches are
significantly different than VNI/URQMD \cite{vniurqmd}, where the initial stage
of the collision is dominated by a partonic (mostly gluonic) cascade. In the
string-based models, the created charges are typically separated along the beam
axis by a distance characterized by a one fermi scale. It is convenient to
monitor the particles spatial position along the beam axis with the coordinate
$\eta$,
\begin{equation}
\eta=\frac{1}{2}\log\left(\frac{t+z}{t-z}\right),
\end{equation}
where for small distances,
\begin{equation}
\Delta\eta= \Delta z/\tau.
\end{equation}
In string models, as well as in boost-invariant hydrodynamics, $\eta$ is
tightly correlated to the collective rapidity of the surrounding particles,
i.e., $y_{\rm collective}\approx \eta$. If a pair is created at $\tau\sim 1$
fm/c, and is separated by nearly a fm, the separation in $\eta$ at this time is
approximately one unit. Thus, the two particles reside in regions which have a
collective rapidity difference of approximately one unit and they will be swept
apart from one another by the collective flow. Re-interaction only provides a
diffusive contribution which further broadens the separation in $\eta$.

\begin{figure}
\centerline{\includegraphics[width=3.8in]{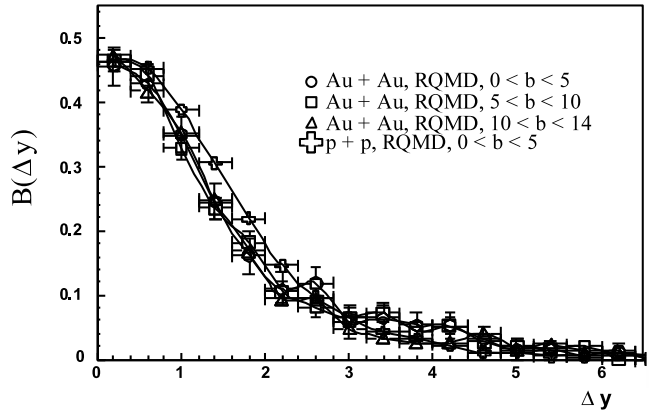}}
\caption{\label{fig:rqmd}
$\pi^+\pi^-$ balance functions for RQMD are shown for both $pp$ and Au+Au
collisions assuming a perfect detector. In contrast to the experimental results
of \cite{starbalance}, the balance function is slightly broader for central Au+Au 
collisions. 
}
\end{figure}

Figure \ref{fig:rqmd} displays $\pi^+\pi^-$ balance functions from RQMD. Due to
a lack of statistics, calculations were performed assuming a perfect
detector. Results are shown for both $pp$ collisions and for Au + Au
collisions. Although the statistics are marginal, the balance function does
appear a few percent wider for Au + Au. This is in stark contrast to the data
which show the balance function narrowing by approximately 15\%. The results of
Fig. \ref{fig:rqmd} are summarized in Table \ref{table:widths}. The half-widths
were calculated by fitting the balance functions to a Gaussian form, similarly
as to what was done in \cite{starbalance}. Results are also shown for HIJING
and HIJING $B/\bar{B}$ \cite{hijingbbar} which incorporate different string
prescriptions than RQMD and ignore rescattering. As shown in \cite{starbalance}
the widths from HIJING do not change from Au+Au to $pp$ as expected.

\begin{table*}
\caption{
Half-widths of balance functions calculated with RQMD, HIJING and HIJING
$B\bar{B}$. Experimental acceptance was not taken into account for these
calculations.}
\label{table:widths}
\begin{tabular}{|c|c|c|c|}
\hline
MODEL & b(fm) & half widths & $\chi^2/$n.d.f.\\ \hline
RQMD(AuAu) & 0-3   & $1.56\pm 0.11$ & 1.16 \\ \hline
RQMD(AuAu) & 3-5   & $1.42\pm 0.11$ & 1.1 \\ \hline
RQMD(AuAu) & 5-7   & $1.41\pm 0.11$ & 2.0 \\ \hline
RQMD(AuAu) & 7-10  & $1.39\pm 0.11$ & 2.9 \\ \hline
RQMD(AuAu) & 10-14 & $1.32\pm 0.11$ & 1.0 \\ \hline
RQMD(pp) &         & $1.48\pm 0.11$ & 5.49 \\ \hline
HIJING(AuAu) & 0-3 & $1.14\pm 0.11$ & 0.92 \\ \hline
HIJ/$\bar{\rm B}$(AuAu) & 0-3 & $1.24\pm 0.08$ & 0.55 \\ \hline
HIJING(pp) &  & $1.18\pm 0.02$ & 5.51 \\ \hline
\end{tabular}
\end{table*}

Table \ref{table:widths} also provides the $\chi^2$ per degree of freedom from
the Gaussian fit. The discrepancy between the shape of the balance function and
a Gaussian is especially apparent for the calculations for $pp$ collisions
which were performed with a large event sample of 40,000 events. Unfortunately,
similar statistics were not attained for the other calculations.

\begin{figure}
\centerline{\includegraphics[width=3.5in]{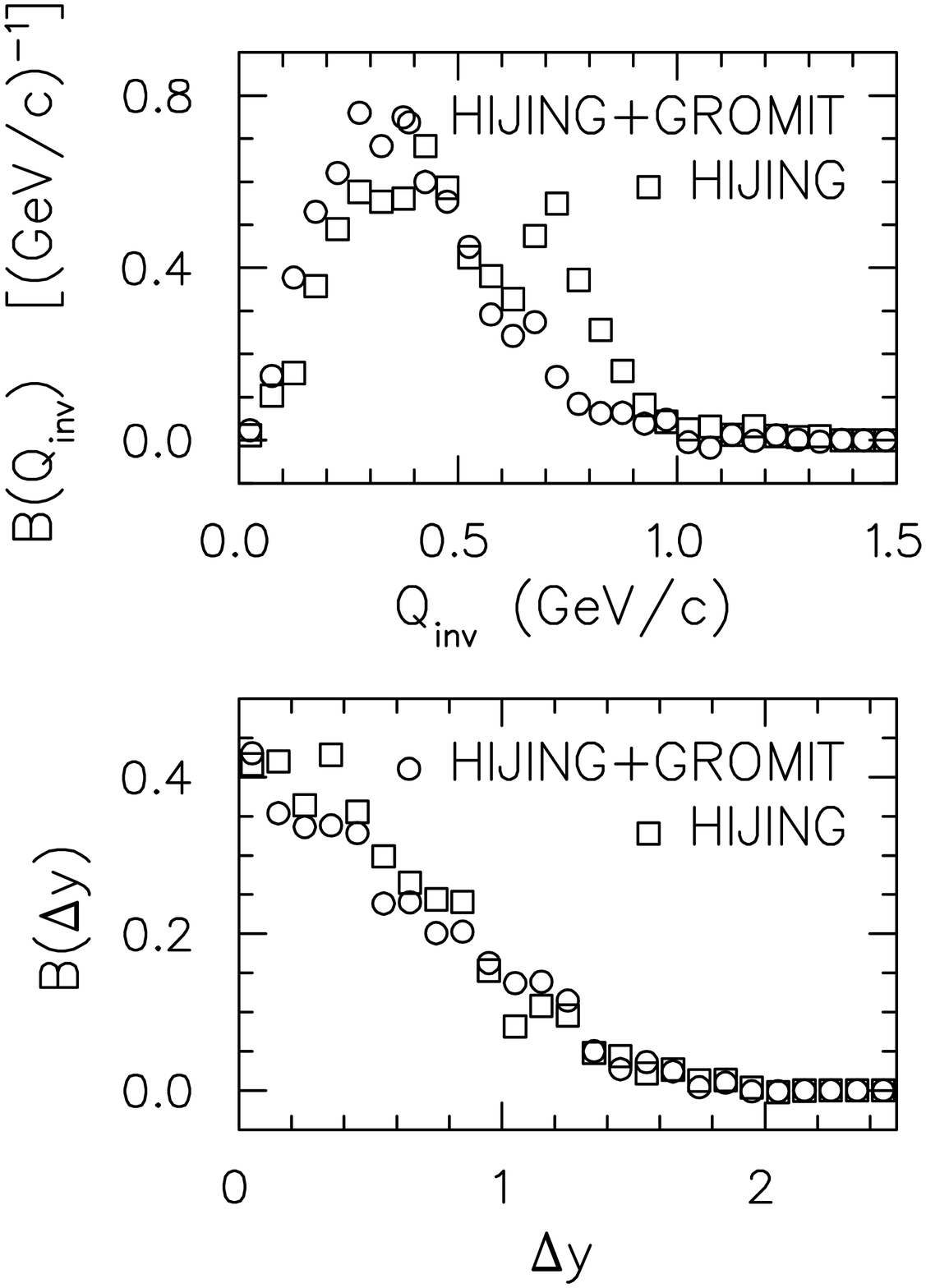}}
\caption{\label{fig:balance_gromit}
Balance functions for HIJING with and without the hadronic cascade GROMIT are
shown as a function of the rapidity difference in the upper panel and as a
function of $Q_{\rm inv}$ in the lower panel. These results have incorporated
the STAR acceptance, but not the efficiency. The incorporation of the cascade
leads to a slightly broadened balance function when analyzed as a function of
$\Delta y$ and a narrower balance function when analyzed as a function of
$Q_{\rm inv}$. This apparent contradiction derives from the failure of both
this model and RQMD to correctly describe the dependence of $\langle
p_t\rangle$ as a function of centrality.}
\end{figure}

Similar results were obtained with the HIJING/GROMIT treatment as can be seen
in the lower panel of Fig. \ref{fig:balance_gromit}. Since the statistics are
improved relative to the RQMD calculation, a simplified version of the STAR
acceptance was taken into account. Pseudo rapidities were confined to
mid-rapidity, $|\eta|<1.1$, transverse momenta were required to be greater than
100 MeV/c and the magnitude of the momenta was confined to be less than 700
MeV/c. In this case, both curves represent Au +Au collisions, but in one case
the re-interaction has been ignored. As with RQMD, the inclusion of hadronic
scatterings appears to marginally broaden the balance function.

The upper panel of Fig. \ref{fig:balance_gromit} shows analogous results from
GROMIT/HIJING binned in $Q_{\rm inv}$ rather than the relative rapidity. Here,
the balance function is slightly narrower after the inclusion of
re-interactions. The difference of the $Q_{\rm inv}$ and $\Delta y$ binnings
derives from two effects. First, during the re-interaction stage, the matter
cools, which reduces the thermal contribution to the width of the balance
function. The cooling affects all three dimensions of relative
momenta. Secondly, the mean $p_t$ decreases due to loss of transverse energy
that accompanies the longitudinal expansion. Since the relative rapidity is
related to the relative momentum via $Q_{\rm long}\sim m_T\Delta y$, where
$m_T$ is the transverse mass, a reduction of the average transverse mass will
result in a narrower balance function when plotted as a function of $Q$ when
the width of the balance function in relative rapidity is unchanged. Thus, part
of the discrepancy between GROMIT/HIJING and data can be attributed to the
failure to describe the behavior of the mean transverse mass as a function of
centrality, which increases by $\sim$ 10\% for central collisions in
experiment. As described in
\cite{balancedistortion}, failure caused by a model not fitting the $p_t$
spectra vs. failure stemming from not describing the dynamics of charge balance
can be better distinguished by analyzing the balance functions in $Q_{\rm
inv}$, or better yet, $Q_{\rm out}$, $Q_{\rm side}$ and $Q_{\rm long}$.

Another striking aspect of Fig. \ref{fig:balance_gromit} is the disappearance
of the $\rho$ peak for $Q_{\rm inv}$ near 700 MeV/c. This is also a natural
consequence of cooling which lowers the $\rho/\pi$ ratio as the temperature
falls to near 100 MeV near breakup.

\section{The Canonical Blast Wave Model}
\label{sec:thermal}

One of the most notable results from the first two years of RHIC data is the
success of the blast-wave model in describing spectra, particle yields and
correlations. The blast-wave model provides an especially important benchmark
for balance functions. Assuming that emission from the final state has a
thermal character, the narrowest balance function would constrain charges and
their balancing partners to be emitted from identical space-time regions. The
width of the balance function would then be determined principally by the
breakup temperature, and to a lesser degree by the resonance composition
\cite{bozek}. If the evolution of the reaction was characterized by an
expanding charge-less long-lived gluonic mire which hadronized close to the
breakup, it would be natural to confine particles and their balancing charges
to be emitted in close proximity. Relaxing this constraint by incorporating
non-zero correlation lengths can only broaden the balance function
\cite{balancedistortion}.

Since charge conservation is local, and since the charge does not have time to
mix throughout the entire collision volume, one needs to choose a volume for
performing the canonical calculations. This scale should be determined by the
distance charge might diffuse by the time the system has reached chemical
freeze-out. We refer to each of these sub-volumes as a domain and view the
entire system as a collection of independent domains.

Of the numerous variations of the blast wave model
\cite{tomasik,tomasikwiedemannheinz,siemensrasmussen,pratt84,retiere}, we
employ a simplified version of this model, where emission occurs at a single
proper time from thermal sources moving with a linear velocity profile.  The
model is characterized by a kinetic freeze-out temperature $T_k$, a chemical
freeze-out temperature $T_\mu$, the velocity at the edge $v_{\rm max}$ and the
maximum radius $R$. The extent along the beam direction is assumed to be
infinite and the distribution of domains is assumed to be uniform in the
transverse direction.
\begin{equation}
\frac{dN}{rdr}=\left\{
\begin{array}{c c}
{\rm constant}, ~r<R\\
0, ~r>R
\end{array}\right.
\end{equation}
The transverse rapidity of domains, $y_{t,domain}$, is assumed to follow the
position $r$.
\begin{equation}
y_{t,domain}=y_{t,{\rm max}}\frac{r}{R}.
\end{equation}

The longitudinal rapidity of the domain, $y_{\rm domain}$ is chosen randomly
within $\pm 3$ units of zero. Since thermal rapidities are on the order of one
to two units, this effectively represents a boost-invariant source.  For this
study, we assume that every particle in a given domain is emitted from a
thermal source moving with the velocity of that domain. If the domain has a
significant spatial extent in the beam direction, the source rapidities would
be spread around the domain rapidity by a finite amount which would broaden
$B(\Delta y)$ \cite{balancedistortion}. Thus, the balance functions shown here
represent the narrowest possible result for a thermal model. The only way to
provide further significant narrowing would be to lower the breakup
temperature, which would require a corresponding increase in the transverse
collective flow.

The simplest model of each source would be that of particle/anti-particle
pairs, i.e., each $\pi^+$ would be balanced by a $\pi^-$ and each $K^+$ would
be balanced by a $K^-$. However, charge mixing and chemical equilibration
spreads the balancing of charge among multiple constituents. For instance the
electric charge and strangeness of a $K^+$ might be balanced by a $\bar{p}$ and
a $\Lambda$. To account for these effects, we generate all the particles from a
source in a manner which is consistent with the canonical ensemble. The
chemical composition of each source will be governed by a volume $V_\mu$, a
temperature $T_\mu$, and the constraint of zero net electric charge,
strangeness, isospin projection $I_3$, and baryon number. After generating the
particles with $T_\mu=175$ MeV, the particles will have their momenta
reassigned according to the blast-wave prescription mentioned above with a
break-up temperature, $T_k=120$ MeV, and a maximum transverse flow velocity of
$0.7c$. The particles are given coordinates corresponding to the position of
the domain and are then boosted by the domain velocity. Mesons were chosen from
the flavor octet and singlet ground state pseudoscalars and pseudovectors,
while baryons were chosen from the ground state baryon decuplet and octet. The
particles were then decayed according to measured decay rates and branching
ratios\cite{pdg}.

In order to perform the Monte Carlo generation of particles from a given
thermal source, one must first calculate the canonical partition functions for
a fixed charge ${\vec Q}$ and a fixed number $A$. This is accomplished by
employing recursion relations, which have been applied to a variety of problems
in nuclear statistical physics where charge and symmetry constraints play an
important role. These applications include multifragmentation
\cite{mekjian,dasguptamekjian,prattdasgupta}, nuclear level densities
\cite{prattfermi}, isospin distributions for pions \cite{senisospin} and a
parton gas confined to a color singlet \cite{finiteqgp}. For the calculations
presented here, Bose and Fermi effects will be neglected and only additive
charges will be considered. This results in a straight-forward recursion
relation for the partition function.
\begin{equation}
Z_{A,{\vec Q}}(T,V)=\sum_k \frac{a_k}{A}\omega_k(T,V) 
Z_{A-a_k,{\vec Q}-{\vec q}_k}(T,V),
\end{equation}
where $k$ labels the particular species, e.g., $\pi^+$, $\pi^0$, $p$,
$\Delta^{++}$. The partition function of a single particle of type $k$ is
$\omega_k$. The number $A$ counts the sum, $\sum_k a_k$, where $a_k$ can be any
positive integer. Since the method will consider all $A$, the choice of $a_k$
is arbitrary. We choose $a_k$ to be unity for all stable particles and to equal
two for unstable particles.

Generating a set of particles from one domain is performed with the following
steps,
\begin{enumerate}
\item Generate the partition function, $Z_{A,{\vec Q}}(T,V)$ for all $A$ and 
${\vec Q}$.
\item Choose a number of hadrons, $A$, proportional to the weight 
$Z_{A,{\vec Q}=0}$.
\item Choose a species $k$ proportional to the weight 
$\omega_k Z_{A-1,{\vec Q}-{\vec q}_k}/Z_{A,{\vec Q}}$.
\item Return to step 2 after replacing, $A\rightarrow A-a_k$, ${\vec
Q}\rightarrow {\vec Q}-{\vec q}_k$.
\end{enumerate}
This procedure naturally ends when $A={\vec Q}=0$. Since particles from one
domain are uncorrelated with particles from another domain, the balance
function need only sum over pairs from the same domain.

To prove that this prescription is consistent with the partition function, one
must show that there is an equivalence between the partition function and a sum
over weights of independent ordered paths. Each path ${\cal P}(A,{\vec
Q};n,k_1\cdots k_n)$ is an ordered list of $n$ particles, $k_1\cdots k_n$ which
sums to the correct $A$ and $\vec{Q}$.  One can then assign a weight for each
path,
\begin{equation}
w(A,{\vec Q};n,k_1\cdots k_n)\equiv \prod_{i=1}^n
\omega_{k_n}\frac{a_{k_n}}{\sum_{i\le n} a_{k_i}}.
\end{equation}
One can then define $\Omega(A,{\bf Q})$ as the sum over such paths that yield
$A$ and $\vec{Q}$. 
\begin{equation}
\Omega(A,\vec{Q})\equiv \sum_{n,k_1\cdots k_n,
	 s.t. \sum a_{k_i}=A,\sum {\vec{q}_{k_i}}=\vec{Q}}
w(A,\vec{Q},k_1\cdots k_n).
\end{equation}
Since the contribution from each path is a product, and since the last term can
be written in terms of $A$, ${\vec Q}$ and the properties of $k_n$, one can
write a recursion relation for $\Omega(A,\vec{Q})$ by factoring the last term,
\begin{equation}
\Omega(A,\vec{Q})=\sum_{k_n}
\frac{a_{k_n}}{A}\omega_{k_n}\Omega(A-a_k,\vec{Q}-\vec{q}_k,k_1\cdots k_n),
\end{equation}
which is the same recursion relation used for the partition
function. Furthermore, since $\Omega(A=0)=Z(A=0)=1$, $\Omega$ and the partition
function $Z$ are identical. Once the partition function can be identified as a
sum over weights of independent paths, one can justify the Monte Carlo
procedure outlined above.

Since charge conservation is enforced on a domain-by-domain basis, there are no
inter-domain contributions to the balance function from this procedure. This
greatly accelerates calculation as one need only consider pairs from within the
same domain when calculating the distributions $N_{+-}\cdots$ used to
construct the balance function.

The resulting balance functions are only sensitive to the choice of domain
volume if the volumes are near or below a few dozen fm$^3$. For the
calculations presented later in this paper, the volume was chosen to be 64
fm$^3$. This would imply, that at the point where chemical freeze out occurs,
charge conservation is enforced on a length scale of $\sim 4$ fm. If one were
to choose a much smaller volume, the likelihood that the charge of a $\pi^+$
would be balanced by a $\pi^-$, vs. the likelihood of being balanced by an
assortment of other particles, would be increased by several percent. This
would result in a larger normalization for the $\pi^+\pi^-$ balance function.

\section{Interdomain correlations}
\label{sec:interdomain}

The philosophy of the balance function is predicated on the assumption that the
background subtraction in the balance function's numerator will statistically
isolate balancing pairs. Of course, one must consider the spread of the
balancing charge among other hadrons in the same domain. This was taken into
account with the Monte Carlo procedure outlined in the previous section. This
procedure accounts for strong interactions between neighbors, but only those
resonant interactions included in the list of particles. For example, the
interaction $\pi^+\pi^-\leftrightarrow \rho^0$ would be included by
incorporating the $\rho^0$ into the resonance list. Given that resonances alter
the shape of the balance function at the 10\% level, and that two-particle
phase shifts are largely driven by resonant interactions, this procedure should
crudely account for the strong interaction between neighbors.

Unfortunately, the procedure thus far does not account for the Coulomb
interaction which was shown to be non-negligible in
\cite{balancedistortion}. The method also ignores non-resonant strong
interactions between neighbors and correlations from identical-particle
interference. The long-range Coulomb interaction is especially important as it
extends beyond the particle's neighbors to those particles generated in
separate domains. A charged particle will effectively polarize other pairs
increasing the likelihood that a particle of opposite charge is emitted at
smaller relative momentum than the balancing charge. This results in an
enhancement to the balance function at relative momenta less than a few hundred
MeV/c and a suppression at larger relative momentum. Non-resonant strong
interactions can also affect the balance function since they provide a
correlation between like-sign pairs that is different from the correlation they
induce between unlike-sign pairs. Finally, identical particle interference can
affect balance functions by increasing the probability that two charges of the
same sign have small relative momentum. This results in a hole in the balance
function for relative momenta below $\sim 100$ MeV/c. However, the Coulomb
again dominates at very low relative momentum and a sharp peak is expected for
relative momenta near or below 10 MeV/c. Since the Coulomb interaction is long
range, interdomain correlations from Coulomb interactions increase with the
centrality of the collisions because a given particle will be correlated with
an increasing number of other particles as the multiplicity of the collision is
increased. Correlations from short-range interactions should be less sensitive
to the multiplicity since the number of neighbors depends principally on the
breakup density rather than the system size. The distortion from
identical-boson interference will be confined to a decreasing range of relative
momentum as the system size increases. However, since there is an increasing
number of particles per element of momentum, the magnitude of the distortion
increases with centrality while the range of the distortion shrinks.

In \cite{balancedistortion} the interdomain correlation was modeled by
generating two pairs of particles, $a_1,a_2$ and $b_1,b_2$. The particles $a_1$
and $a_2$ were antiparticle reflections of one another, as were the particles
$b_1$ and $b_2$. In the limit of small domains, where the multiplicity of a
domain is never more than two, this is equivalent to the present problem. The
effects of the interdomain Coulomb force was included by calculating a
correlation weight formed by the product of all two-particle interdomain
correlation functions, $W_{AB}=C_{a1,b1}(p_{a1},p_{b1})C_{a1,b2}(p_{a1},p_{b2})
C_{a2,b1}(p_{a2},p_{b1})C_{a2,b2}(p_{a2},p_{b2})$. Each two-particle
correlation function was generated from a simple Gaussian source functions. For
each set of pairs the weight was applied to all the distributions, $N_{++},
N_{+-} \cdots$, used to calculated the balance function numerators as well as
the one-particle distributions, $N_+$ and $N_-$ used to form the
balance-function denominators. For each $ab$ pair, the number of $cd$ pairs
sampled was chosen to achieve consistency with the experimental multiplicity.

The method of \cite{balancedistortion} has two shortcomings. First, the product
ignored correlations of correlations. For instance, the product $C_{ac}C_{ad}$
neglects the fact that if $a$ and $c$ are correlated, than they are more likely
to have been close to one another in coordinate space, therefore $a$ and $d$
would more likely be close to one another since $c$ and $d$ are from the same
domain. The more obvious limitation of this approach is that it can not handle
the chance that the domain consists of several dozen particles, with the
constraint of charge conservation being spread throughout the domain.

The first shortcoming can be accounted for by replacing the product of
correlation functions with a product of squared wave functions,
\begin{equation}
\label{eq:WABfull}
W_{AB}= \left\langle 
\prod_{a\in A,b\in B}|\phi_{ab}(p_a-p_b,x_a-x_b)|^2
\right\rangle.
\end{equation}
Here, $\phi_{ab}$ is the relative wave function for an outgoing plane wave with
asymptotic relative momentum, $p_a-p_b$, and the averaging covers the
distribution of source points $x_a$ and $x_b$. If the expectation of the
products of the squared wave function were replaced by the product of
expectations, Eq. (\ref{eq:WABfull}) would become the product of correlation
functions, and would be equivalent to the method of \cite{balancedistortion} in
the case where each domain had two particles. 

The pairs were generated according to the blast wave prescription described in
the previous section, with both particles from a given pair being assigned the
same point in coordinate space. Balance-function numerators were then
calculated using only the two pairs, and ignoring the contribution where both
particles originate from the same pair. Rather than incrementing a bin by unity
when an appropriate pair is found, the quantities $N_{++}$ and $N_{+-}$ were
incremented by $W_{AB}$. Since the source of the pairs was confined to a region
of $-\eta_{\rm max}<\eta<\eta_{\rm max}$ where $\eta_{\rm max}=2$, the balance
function is then scaled upwards by a factor of the number of pion pairs which
should be emitted within the central four units of rapidity. For the
calculations in this paper it was assumed that the number of pairs would be 200
per unit rapidity.

For a perfect detector, the weight $W_{AB}$ is added to both $N_{++}$ and
$N_{+-}$. Thus, the normalization for the balance function is unchanged by
inter-domain correlations, i.e., the inter-pair interactions effectively
polarize the pairs but they do not enhance the overall number of one charge
vs. another. This constraint is relaxed after detector acceptance is taken into
account. 

We believe that this estimate of the distortion from interpair correlations can
only be trusted at the 20\% level. First, some of the pions are emitted
from long-lived resonances and should not contribute. Accounting for these
effect should reduce the effect by several tens of percent. Other aspects of
the approximation, such as neglecting interactions with other particles and
requiring both particles from a given pair to originate from the same point,
may also affect the answer at the 10\% level.

Wave functions for a pair $ab$ were calculated using full Coulomb wave
functions. As only $\pi^+\pi^-$ correlations are being considered here, the
wave functions were only symmetrized if $a$ and $b$ were identical
pions. Strong interaction corrections were also only applied if $a$ and $b$
were pions. The wave functions were modified by the strong interaction using
the methods reference \cite{corrtail}. In \cite{corrtail} the wave functions
for $|{\bf x}_1-{\bf x}_2|>1$ fm were calculated using phase shifted partial
waves, while effective squared wave functions were applied for small
distances. These effective forms were also completely determined by the phase
shifts. The phase shifts for $\pi^+\pi^-$ were taken from experiment whenever
possible. Unfortunately, they are not well understood for invariant masses
above the two-kaon threshold. 

The dominant phase shift is in the $\ell=1,I=1$ channel which is driven by the
$\rho$ resonance. A Breit-Wigner function was applied for this channel.
\begin{eqnarray}
\label{eq:rhoofdelta}
\tan\delta&=&\frac{\Pi_I}{M_0^2-M^2},\\
\Pi_I&=&\Pi_{I,0}\frac{M_0}{M}\left(\frac{q}{q_0}\right)^3F(q,q_0),\\
\Pi_{I,0}&=&\Gamma_0 M_0.\\
\end{eqnarray}

The $\ell=0,I=0$ channel also plays an important role in affecting the wave
function. Although phase shift analyses do not reveal a sharp peak as in a
resonance \cite{pichowsky, kaminski, grayer,rosselet,shrinivasan}, the phase
shifts are considerable, rising steadily from zero at threshold to
approximately 90 degrees at $M=2M_K\sim 1$ GeV, where the kaon channel
opens. At the two-kaon threshold, the behavior of the phase shifts becomes
complicated and an inelastic treatment becomes warranted. Since one uses
derivatives of the phase shifts to find the density of states, interpolating
data for phase shifts can be dangerous due to noise in the experimentally
determined phase shifts. Thus, we apply a simple form that describes the
general behavior,
\begin{eqnarray}
\tan(\delta_{I=0,\ell=0})&=&\frac{aq}{1-(q/a_2)+(q^2/a_3^2)},\\
a&=&0.204/m_\pi, a_2=290 ~{\text MeV/c}, a_3=625 ~{\text MeV/c}.
\end{eqnarray}
The coefficient $a$ is chosen to reproduce the scattering length, which is
small due to constraints from chiral symmetry \cite{kermani,pichowsky}. The
remainder of the denominator is chosen to provide crude agreement with measured
phase shifts up to invariant masses of 1 GeV.  Since the phase shift rises only
half as high as the $I=1,\ell=0$ phase shift, and since it contributes only one
third as much from a lower spin degeneracy, this channel has a fairly marginal
impact on the wavefunctions.

Other phase shifts also contribute: $(I=2,\ell=0)$, $(I=0,\ell=2)$ and
$(I=2,\ell=2)$. Since none of these phase shifts exceed more than a few
degrees, they make nearly negligible contributions to the density of
states. For the $(I=2,\ell=0)$ channel, we apply an effective range expansion
\cite{losty},
\begin{eqnarray}
\tan\delta=\frac{qa}{1+aq^2R_{\rm eff}/2},\\
a=-0.13 ({\text MeV/c})^{-1}, R_{\text eff}=1.0 {\text fm}.
\end{eqnarray}
For the $(I=0,\ell=2)$ phase shift, the data \cite{estabrooks} are rough, and
we make a simple expansion,
\begin{equation}
 \tan(\delta_{I=0,\ell=2})=ax^5,~~x=\frac{q}{1+q^2/\Lambda^2},
\end{equation}
where $a=6.2$ GeV$^{-1}$, $\Lambda=1$ GeV/c. For the $(I=2,\ell=2)$ partial
wave, we use the same expansion \cite{losty} with the parameters $a=8.4$
GeV$^{-1}$ and $\Lambda=0.4$ GeV/c. Although none of the these last three
channels are particularly well understood, it is clear they do not have a
substantial impact on the result.

\section{Canonical Blast Wave Results}
\label{sec:thermalresults}

Calculations for the canonical blast-wave calculations described in
Sec. \ref{sec:thermal} and Sec. \ref{sec:interdomain} were performed assuming a
temperature of $T_k= $120 MeV and a maximum transverse velocity of 0.7$c$ to
provide crude agreement with spectra. The transverse radius was chosen to equal
13 fm and the breakup time $\tau$ was chosen to be 10 fm/c to reproduce HBT
results. The chemical composition was governed by a temperature $T_\mu=175$ MeV
as suggested by fits of the particle yields \cite{magestro}.

\begin{figure}
\centerline{\includegraphics[width=3.5in]{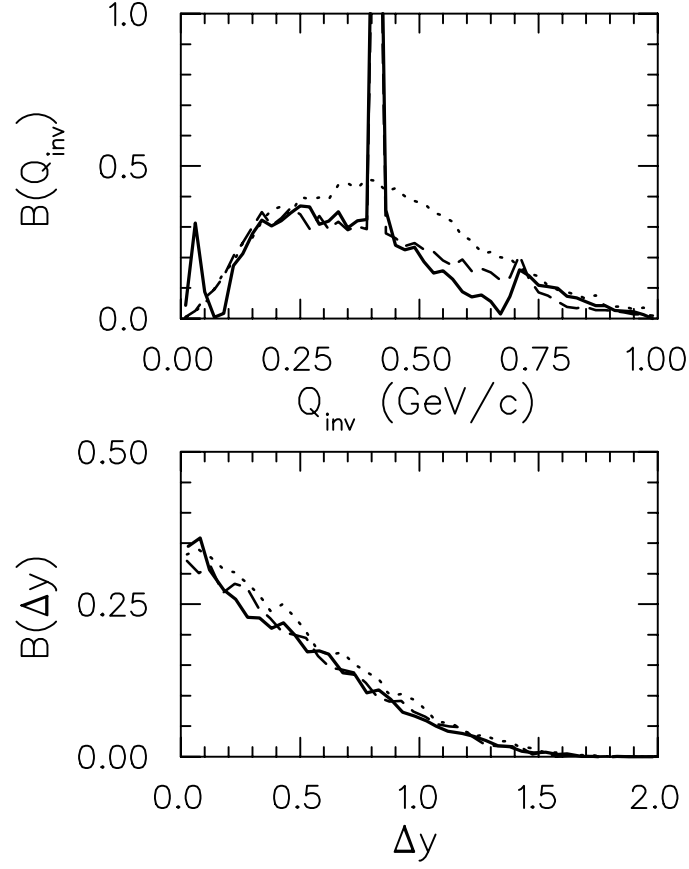}}
\caption{\label{fig:balance_Tk120_Tmu175}
Balance function results for a pion gas (dotted lines) and for a resonance gas
(dashed lines) are shown for the canonical thermal blast-wave model. Resonances
clearly narrow the balance function when plotted against $Q_{\rm inv}$ (upper
panel) but have little effect on the balance function when plotted as a
function of the rapidity difference (lower panel). Balance functions corrected
for interpair correlations (solid lines) again differ when plotted as a
function of $Q_{\rm inv}$. The additional distortion arises from the effects of
Coulomb, symmetrization and a more sophisticated treatment of the strong
interaction.}
\end{figure}

Figure \ref{fig:balance_Tk120_Tmu175} displays results for balance functions
analyzed in both $\Delta y$ and $Q_{\rm inv}$. The acceptance and efficiency of
the STAR detector have been applied to the calculation of the distributions
that are used to construct the balance function. These accepted particles
include products of weak decays of strange baryons and $K_s$ mesons. Results
are shown both with and without the contribution from inter-domain
correlations. Since the interaction of $\pi^+\pi^-$ pairs through the $\rho^0$
meson was incorporated into the relative wave functions, the contribution from
$\rho^0$s were eliminated before the inter-domain correlations were added.

The relative contribution of resonances and strong interactions can be
estimated by considering the chance that a charged pion originated from a
resonance that also produced a pion of the opposite charge. Although such
estimates suggest contributions on the level of 10 to 20\%, they seem less
visible when viewing the results. This is because the separation of a
$\pi^+\pi^-$ pair in momentum is not much different coming from a resonance as
compared to a distribution governed by two Boltzmann terms. In fact, the
largest inter-domain effect derives from the Coulomb interaction which moves
some of the strength of the balance function to smaller relative momentum.

\begin{figure}
\centerline{\includegraphics[width=3.5in]{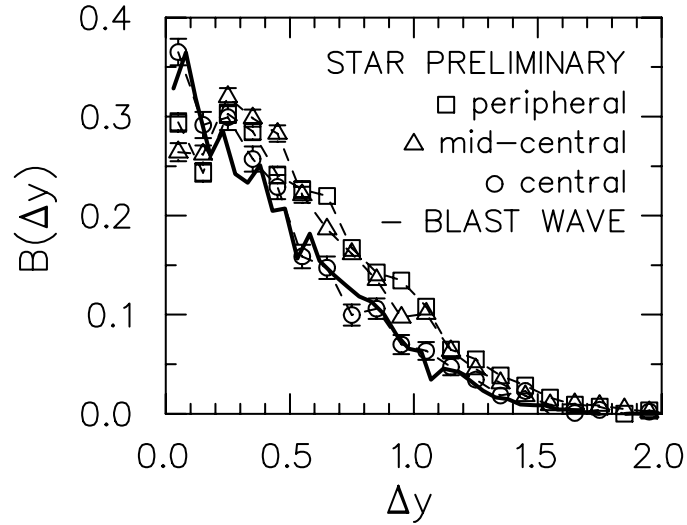}}
\caption{\label{fig:starcomparison}
Balance functions from 200$A$ GeV Au+Au collisions measured by STAR are
compared to the canonical blast-wave model described in the text. The model
should set a lower bound for the width of a balance function provided the
particles are emitted thermally. The remarkable agreement with the data
suggests that charge conservation remains highly localized at breakup.}
\end{figure}

An absolute comparison with STAR results is displayed in
Fig. \ref{fig:starcomparison}. Both the height and width of the correlation are
remarkably well matched by the thermal model. Since the canonical blast-wave
calculation assumed the balancing charges were always emitted from sources
moving with the same velocities, this represents the narrowest possible balance
function one can generate with a final breakup temperature of 120 MeV. This
appears to corroborate the scenario of late-stage hadronization, i.e., the
existence of the quark-gluon plasma.

Before claiming that Fig. \ref{fig:starcomparison} provides ``proof'' of the
delayed hadronization, it is important to list a few qualifiers. First, the
breakup temperature might be lower than the 120 MeV assumed here. Some
estimates of the breakup temperature are near 100 MeV \cite{retiere}, though
after the inclusion of resonances, the temperatures are usually closer to 120
MeV. It should be kept in mind that an anomalously high yield of light neutral
hadrons, e.g., $\eta$ mesons, could result in a narrower balance functions. A
lower temperature or high $\eta$ yield would narrow the balance function and
permit the incorporation of a finite spread in the domain size along the beam
axis. In reference \cite{balancedistortion} it was shown that since the thermal
and diffusive contributions add in quadrature, the diffusive contribution is
negligible unless the charge separates in coordinate space by of the order of
one half unit of $\Delta\eta=\Delta z/\tau$ or more.

\section{Summary}
\label{sec:summary} 
The observed narrowing of the balance function is both qualitatively and
quantitatively consistent with the scenario of delayed hadronization, which
should result in the separation of balancing charges being determined primarily
by the collective flow and temperature of the breakup stage. The experimental
results were well fitted by a canonical blast-wave calculation assuming a
temperature of 120 MeV and a maximum transverse collective velocity of
$0.7c$. By lowering the breakup temperature and raising the collective flow in
the model, one should still be able to describe the pion spectra while
producing a balance function that is even narrower than that observed
experimentally. This would then allow one to accommodate a significant
longitudinal size for the charge-conservation domains in the thermal
model. This ambiguity would be clarified by a multi-dimensional analysis of the
balance function in the relative momenta of the pions. If the charge production
is indeed delayed, the width of the balance function should be the same for
$Q_{\rm out}$, $Q_{\rm side}$ and $Q_{\rm long}$ \cite{balancedistortion}.

Furthermore, the observations are opposite to the behavior predicted by purely
hadronic models which predict a modest broadening of the balance function as
centrality is increased. Unfortunately, the width of the balance function in
relative rapidity is significantly affected by both the final temperature and
the collective flow of the matter. Since HIJING/GROMIT gives the wrong behavior
for the mean $p_t$ as centrality increases, some of the failure to describe the
centrality dependence of the balance function might derive from the inability
to reproduce spectra rather than an inability to correctly describe the
mechanism for charge production. This ambiguity would also be resolved by a
multi-dimensional analysis.

The measurement, analysis and phenomenology of charge balance functions is in
its nascent stage. Nonetheless, the first analysis of RHIC data already provide
important limits for understanding the production and separation of balancing
charges in central gold collisions. There is tremendous promise for this class
of observable as it is extended to more dimensions and to more species, e.g.,
$\bar{p}p$ and $K^+K^-$. These more detailed analyses should then provide a
fingerprint for making unambiguous statements concerning the production and
dissipation of charge in collisions at RHIC. Since delayed hadronization is
synonymous with delayed production of charge, these measurements can address
the fundamental question of whether a new phase of matter has been created at
RHIC.

\begin{acknowledgments}
The comparison with STAR data could not have been performed without the
acceptance and efficiency codes generously provided by Manuel Calderon. This
work was supported the Natural Sciences and Engineering Research Council of
Canada, le Fonds Nature et Technologies of Qu\'ebec, by the U.S. National
Science Foundation, Grant No.  PHY-02-45009, and by the U.S. Department of
Energy, Grant No.s DE-FG02-03ER41259 and DE-AC02-98CH10886,
\end{acknowledgments}


\begin{thebibliography}{99}
\bibitem{bassdanpratt} S.A. Bass, P. Danielewicz, and S. Pratt,
Phys. Rev. Lett. {\bf 85},  2689 (2000).
\bibitem{starbalance} J. Adams et al. Phys. Rev. Lett. {\bf 90}, 172301 (2003).
\bibitem{rqmd} RQMD, H. Sorge, H. St\"{o}cker and W. Greiner, Ann. Phys. {\bf
192}, 266 (1989).
\bibitem{hijing} X.N. Wang and M. Gyulassy, Comput. Phys. Commun. {\bf 83}, 307
(1994).
\bibitem{gromit} S. Cheng, S. Pratt, P. Csizmadia, Y. Nara, D. Molnar, 
M. Gyulassy, S.E. Vance and B. Zhang, Phys. Rev. C{\bf 65}, 024901 (2002).
\bibitem{balancedistortion} S. Pratt and S. Cheng, Phys.Rev. C {\bf 68}, 
014907 (2003).
\bibitem{rhicbjorkenestimate} K. Adcox et al., Phys. Rev. Lett. {\bf 87},
052301 (2001).
\bibitem{mcgillchargefluc} Q.H. Zhang, V. Topor Pop, S. Jeon and C. Gale ,
Phys. Rev. C{\bf 66}, 014909 (2002).
\bibitem{jeonpratt} S. Jeon and S. Pratt, Phys. Rev. C{\bf 65}, 044902 (2002).
\bibitem{venus} VENUS, K.~Werner, Phys.~Rep.~{\bf 232}, 87(1993).
\bibitem{hsd} HSD, W. Cassing and S. Juchem, Nucl. Phys. A {\bf 672}, 417 (2000).
\bibitem{art} ART, B.~A.~Li and C.~M.~Ko, Phys.~Rev.~{\bf C52}, 2037 (1995).
\bibitem{nexus} NEXUS, H.J. Drescher, M. Hladik, K. Werner, and S. Ostapchenko,
Nucl. Phys. Proc. Suppl. A{\bf 75}, 275 (1999).
\bibitem{hijingbbar} S.E. Vance and M. Gyulassy, Phys. Rev. Lett. {\bf 83},
1735 (1999).
\bibitem{urqmd} URQMD, S.~A.~Bass, M.~Belkacem, M.~Bleicher, M.~Brandstetter,
L.~Bravina, C.~Ernst, L.~Gerland, M.~Hofmann, S.~Hofmann, J.~Konopka, G.~Mao,
L.~Neise, S.~Soff, C.~Spieles, H.~Weber, L.~A.~Winckelmann, H.~St\"ocker,
W.~Greiner, C.~Hartnack, J.~Aichelin and N.~Amelin, Progr. Part. Nucl. Physics
Vol. {\bf 41},~225~(1998).
\bibitem{ampt} AMPT, B. Zhang, C.M. Ko, B.A. Li, Z. Lin, Phys. Rev. C {\bf 61},
067901 (2000).
\bibitem{jpciae} JPCIAE. (string-fragmentation combined with hadronic cascade),
Sa Ben-Hao, Tai An, Wang Hui, and Liu Feng-He, Phys. Rev. C{\bf 59}, 2728
(1999).
\bibitem{qgsm} QGSM, N.S. Amelin et al., E.F. Staubo and L.P. Csernai,
Phys. Rev. D {\bf 46}, 4873 (1992).
\bibitem{vniurqmd} VNI/URQMD, (partonic cascade with hadronic cascade),
S.A.~Bass, M.~Hofmann, M.~Bleicher, L.~Bravina, E.~Zabrodin, H.~St\"ocker and
W.~Greiner, Phys.~Rev.~{\bf C60},~021901~(1999).
\bibitem{bozek} P. Bozek, W. Broniowski and W. Florkowski, preprint, 
www.arXiv.org, nucl-th/0310062 (2003).
\bibitem{tomasik} B. Tom\'asik, To appear in the proceedings of 38th Rencontres
de Moriond on QCD and High-Energy Hadronic Interactions, Les Arcs, Savoie,
France, 22-29 Mar 2003, preprint: www.arXiv.org, nucl-th/0304079.  (2003).
\bibitem{tomasikwiedemannheinz} B. Tom\'asik, U.A. Wiedemann and U. Heinz, 
Heavy Ion Phys. {\bf 17}, 105
(2003).
\bibitem{siemensrasmussen}
P. Siemens and J. Rasmussen, Phys. Rev. Lett. {\bf 42}, 880 (1979).
\bibitem{pratt84} S. Pratt, Phys. Rev. Lett. {\bf 53}, 1219 (1984).
\bibitem{retiere} F. Retiere, Proceedings of the Intl. Workshop on Physics of
the Quark-Gluon Plasma, Palaiseau, France 2001, www.arXiv.org, nucl-ex/0111013.
\bibitem{pdg} Particle Data Group, K. Hagiwara et al., Phys. Rev. D{\bf 66},
010001 (2002).
\bibitem{mekjian}
K.~C. Chase and A.~Z. Mekjian, Phys. Rev. C{\bf 52},  R2339  (1995).
\bibitem{dasguptamekjian}
S. Das~Gupta and A.~Z. Mekjian, Phys. Rev. C{\bf 57},  1361  (1998).
\bibitem{prattdasgupta}
S. Pratt and S. Das~Gupta, Phys. Rev. {\bf C62},  044603  (2000).
\bibitem{prattfermi} S. Pratt, Phys. Rev. Lett. {\bf 84},  4255  (2000).
\bibitem{senisospin} S. Cheng and S. Pratt, Phys. Rev. C {\bf 67}, 044904
(2003).
\bibitem{finiteqgp} S. Pratt and J. Ruppert, Phys. Rev. C {\bf 68}, 02904
(2003). 
\bibitem{corrtail} S. Pratt and S. Petriconi, www.arXiv.org, nucl-th/0305018
  (2003).
\bibitem{pichowsky} M.A. Pichowsky, A. Szczepaniak and J.T. Londergan, 
Phys.Rev. D{\bf 64}, 036009 (2001).
\bibitem{kaminski} R. Kaminiski, L. Lesniak and K. Rybicki,
Acta. Phys. Polon. B{\bf 31}, 895 (2000).
\bibitem{grayer} G. Grayer, et al., Nucl. Phys. B{\bf 75}, 189 (1974).
\bibitem{rosselet} L. Rosselet, et al., Phys. Rev. D{\bf 15}, 574 (1977).
\bibitem{shrinivasan} V. Shrinivasan et al., Phys. Rev. D{\bf 12}, 681 (1975).
\bibitem{kermani} M. Kermani et al., Phys. Rev. C{\bf 58}, 3431 (1998).
\bibitem{losty} M.J. Losty, V. Chaloupka, A. Ferrando, L. Montanet, E. Paul,
D. Yaffe, A. Zieminski, J. Alitti, B. Gandois and J. Louie, 
Nucl. Phys. B{\bf 69}, 185 (1974).
\bibitem{estabrooks} P. Estabrooks and A.D. Martin, Nucl. Phys. B{\bf 79}, 
301 (1974).
\bibitem{magestro} P. Braun-Munzinger, D. Magestro, K. Redlich and J. Stachel,
Phys. Lett. B {\bf 518}, 41 (2001).
\end{thebibliography}
\end{document}